\documentstyle[12pt]{article}

\begin{document}

\begin{titlepage}
\begin{flushright} {IC/94/201} \end{flushright}

\vskip 2.0truecm

\begin{center}
{\Large \bf
Central Extension of a \vskip 3mm
New $W_\infty$-Type Algebra}
\end{center}

\vskip 1.0cm

\begin{center}
M. N. Stoilov$^{\star}$ \\
{\it International Centre for Theoretical Physics, Trieste 34100,}
{\bf Italy}\\
and\\
{\it Institute for Nuclear Research and Nuclear Energy, \\
Boul. Tzarigradsko Chaussee 72, Sofia 1784,}
{\bf Bulgaria}$^{\dag}$
\vskip 5mm
and
\vskip 5mm
R. P. Zaikov$^{\ddag}$ \\
{\it Institute for Nuclear Research and Nuclear Energy, \\
Boul. Tzarigradsko Chaussee 72, Sofia 1784,}
{\bf Bulgaria}
\vskip 1cm
20th July 1994
\end{center}

\vskip 2.0cm

\rm
{\bf Abstract. }
The central extension of a new infinite dimensional algebra
which has both $W_\infty$ and affine $sl(2,R)$ as subalgebras
is found. The critical dimension of the corresponding string
model is $D=5$.
\vfill

\begin{flushleft}
\rule{5.1 in}{.007 in}\\
{\small
$^{\star}$
Supported by Bulgarian National Scientific Found under contract Ph-20-91\\
$^{\dag}$  Permanent Address \\
$^{\ddag}$
Supported by Bulgarian National Scientific Found under contract Ph-318-93}
\end{flushleft}

\end{titlepage}

\newpage

     Recently we obtained \cite{SZ} a new $W_\infty$-type algebra
as a dynamical symmetry algebra  of  the  so  called  `generalized
Chern-Simons  string'.   In  this  letter  our goal is to find the
possible nontrivial central extension of this algebra and the
critical dimension of the model.

 The generalized  Chern-Simons  string  is  characterized  by  the
following  constraint  system
\begin{equation}
P^\mu P_\mu^{(n)} \approx 0, \;\;\;
X^\mu X_\mu^{(n)} \approx 0, \;\;\;
P^\mu X_\mu^{(n)} \approx 0, \;\;\; n=0,1,2,...,
\label{old}
\end{equation}
 where $X_\mu(\tau,\sigma)$ give the embedding of the string into a
$D$-dimensional Min\-kov\-ski spacetime  and   $P_\mu$   are  the
canonically  conjugated  to  $X_\mu$   momenta.    The   notations
$X_\mu^{(n)}$  and  $P_\mu^{(n)}$  used  in (\ref{old}) and bellow
stand for the derivatives
$\partial^n / \partial\sigma^n (X_\mu)$ and
$\partial^n / \partial\sigma^n (P_\mu)$.
Using the basic commutation relations between coordinate and
momentum it is easy to find the equal time commutation relations
between constraints (\ref{old}).

In the present letter instead of (\ref{old}) we shall use another basis
of generators. There are two reasons for this. First, as it is
easy to see, the generators (\ref{old}) are not independent (e.g.
$P^2$ and $PP^{(1)}$). Second, it turns out that
(\ref{old}) is not quite suitable for the estimation of the
central extension. The basis we shall use is
\begin{equation}
{(P^{(n)})^2\over 2}, \;\;\;
{(X^{(n)})^2\over 2}, \;\;\;
P X^{(n)}, \;\;\; n=0,1,2,\dots
\label{new}
\end{equation}
It is convenient to introduce the
functional representation of the above generators
\begin{eqnarray}
{\cal P}_l [f]&=&\int d\sigma f(\sigma)
{(P^{(l)})^2\over 2} \nonumber \\
{\cal X}_m [g]&=&\int d\sigma g(\sigma)
{(X^{(m)})^2\over 2} \label{fg} \\
{\cal G}_n [h]&=&\int d\sigma h(\sigma) P X^{(n)}. \nonumber
\end{eqnarray}
Here $f(\sigma), g(\sigma)$ and $h(\sigma)$
are arbitrary functions on the circle (we consider a closed string
only).

In order to establish the commutation relations in the basis
(\ref{new}) and to find how to pass from this basis to the old one
(the inverse transition from (\ref{old}) to (\ref{new}) is
trivial) one has to learn to express quantities like
${\cal F}_{n,m}[g] = \int d\sigma g(\sigma)
P_\mu^{(n)}P^{\mu (n+m)}$
in terms of
${\cal P}_k [g]$.
Using the identity
$${\cal F}_{n,m}[g] = -{\cal F}_{n,m-1}[g^{(1)}] -
{\cal F}_{n+1,m-2}[g]$$
and the initial condition
${\cal F}_{n,1}[g] = - {\cal P}_n[g^{(1)}]$
one can prove that
\begin{equation}
{\cal F}_{n,m}[g] = \sum_k^{[{m \over 2}]}
(-1)^{m+k}{m \over m-k}{m-k \choose k}{\cal P}_{n+k}[ g^{(m-2k)}],
\;\;\; m>0,\label{chng}
\end{equation}
where $[{m/2}] $ is the greatest integer less than or equal to ${m/2}$.
Using formula (\ref{chng}) and a similar one for
$\int d\sigma f(\sigma) X_\mu^{(n)}X^{\mu (n+m)}$
we obtain the following (equal time) commutation relations between
generators (\ref{new}) (only the nontrivial ones are listed)
\begin{eqnarray}
\left[ {\cal G}_m[f], {\cal G}_n[g]\right] &=&
\sum_{k=0}^n {n\choose k}{\cal G}_{m+k}[f^{n-k}g] -
\sum_{k=0}^m {m\choose k}{\cal G}_{n+k}[fg^{m-k}] \label{px-px} \\
\nonumber\\
\left[ {\cal P}_m[f], {\cal G}_n[g]\right] &=&
\sum_{k=0}^{m+n}\sum_{l=0}^{[{m+k\over 2}]}(-1)^{k+l}
{m+n\choose k}{m+k-l\choose l}{m+k\over m+k-l} \times\nonumber\\
\nonumber\\
&&{\cal P}_l[(gf^{(m+n-k)})^{(m+k-2l)}]  \\
\nonumber\\
\left[ {\cal X}_m[f], {\cal G}_n[g]\right] &=&
\sum_{k=0}^m\sum_{l=0}^{[{a+k\over 2}]}(-1)^{n+k+l+1}
{m\choose k}{a+k-l\choose l} {a+k\over a+k+l}\times\nonumber\\
\nonumber\\
&&{\cal X}_{b+l}[h_{m-k}^{(a+k-2l)}] \label{xx-px} \\
\nonumber\\
\left[ {\cal P}_m[f], {\cal X}_n[g]\right] &=&
(-1)^{m+n}
\sum_{k=0}^{m+n}\sum_{l=0}^m
{m+n\choose k}{m\choose l}  \times\nonumber\\
\nonumber\\
&&{\cal G}_{n+k+l}[(fg^{(m+n-k)})^{(m-l)}] \label{pp-xx}
\end{eqnarray}
In eq.(\ref{xx-px}) $\;\;a=\vert m-n\vert,
\;\;\;b={\rm max}\{m,n\},
\;\;\; h_{m-k} = gf^{(m-k)}\;\;$ for $m\ge n$
and $h_{m-k} = fg^{(m-k)}\;\;$ for $n>m$.

One immediately recognizes two important subalgebras of the above
algebra --- the first one is $sl(2,R)$ with generators
${P^2/2}, {X^2/2}, PX$, and the second is $DOP(S^1)$
with generators $PX^{(n)},\;\;\;n=0,1,2,\dots$.
Both these algebras have unique nontrivial central extensions.
The corresponding two-cocycles are
\begin{equation}
c({\cal G}_m[f],{\cal G}_n[g]) =
c {m! n! \over (m+n+1)!} \int d\sigma f^{(n)} g^{(m+1)}
\label{dop}
\end{equation}
for $DOP(S^1)$ \cite{Feig} and
\begin{equation}
c({\cal G}_0[f],{\cal G}_0[g]) =
- 2 c({\cal P}_0[f],{\cal X}_0[g])
= k \int d\sigma f g^{(1)}
\label{km}
\end{equation}
for $\hat{sl}(2,R)$ \cite{Mieg} (in the Chevalley basis used
here the $sl(2,R)$ Killing metric is proportional to the matrix
$\pmatrix{0&0&1\cr 0&-2&0\cr 1&0&0}$
and this determines the ``index'' structure of (\ref{km})).
It is clear from eqs. (\ref{dop}, \ref{km}) that the constants
$k$ and $c$ have to be equal.
Moreover, a nontrivial two-cocycles
$c({\cal P}_m[f],{\cal X}_n[g])= -c({\cal X}_n[g],{\cal P}_m[f])$
(which generalize the second one in (\ref{km}))
have to exist for any $m$ and $n$ in order for the whole algebra to
possess a central extension.
Using the Jacoby identity
\begin{equation}
c(\left[{\cal G}_l[f],{\cal X}_m[g]\right],{\cal P}_n[h]) +
c(\left[{\cal X}_m[g],{\cal P}_n[h]\right],{\cal G}_l[f]) +
c(\left[{\cal P}_n[h],{\cal G}_l[f]\right],{\cal X}_m[g]) = 0
\label{jac}
\end{equation}
(there are no other nontrivial Jacoby identities except (\ref{jac})
and the one which involves only ${\cal G}_m$ and determines
(\ref{dop})) one can prove that
\begin{equation}
c({\cal P}_m[f],{\cal X}_n[g]) =
-{c\over 2} {(m+n)!(m+n)! \over (2m+2n+1)!}
\int d\sigma f^{(2n)} g^{(2m+1)}
\label{my}
\end{equation}
The cocycles
(\ref{dop}), (\ref{my}) determine the unique nontrivial central
extension of the algebra in consideration.
The following formulas are used to obtain them
\begin{eqnarray}
\sum_{k=0}^n (-1)^k {(m+k)!\over (m+p+k)!} {n\choose k} &=&
{m!(n+p-1)!\over (p-1)!(m+n+p)!},  \hskip 7mm
{\rm for}\; p\ge 1
\label{fe1}\\
\sum_{k=0}^n (-1)^k {(m+p+k)!\over (m+k)!} {n\choose k} &=&
0, \hskip 42mm {\rm for}\; p < n
\label{fe2}\\
\sum_{k=0}^n (-1)^k {(m+p+k)!\over (m+k)!} {n\choose k} &=&
(-1)^n{p!(m+p)!\over (p-n)!(m+n)!}, \;\;\;
{\rm for}\; p\ge n
\label{fe3}\\
\sum_{k=0}^n (-1)^k { p \over  p - k}
{{(m+k)!}^2\over (2m+2k+1)!} {p-k\choose k}& = &
2{m!(m + p)!\over (2m+p+1)!}, \nonumber\\
&&\hskip15mm {\rm for}\;  p=2n-1, 2n, 2n+1
\label{myf}
\end{eqnarray}
The first three of these identities can be found in \cite{Feig}
(see also \cite{Rad}).
One can prove for example (\ref{fe1})
first for $p=1$ using the residue theorem for the function
$1/\left(z \prod_{i=m+1}^{m+n+1} (z-i)\right)$
and then by induction for any $p$.
The second and the third formula ((\ref{fe2}) and (\ref{fe3}))
can be shown analogously.
Formula (\ref{myf}) can be proven by induction following the
chainlet (here $M(n,p)$ denotes identity (\ref{myf}))
\begin{eqnarray}
&&M(0,1)\rightarrow\dots\rightarrow
M(n,2n-1)\rightarrow M(n,2n) \rightarrow \nonumber\\
&&\hskip 3cm \rightarrow M(n,2n+1) \rightarrow
M(n+1,2n+1)\rightarrow\dots \nonumber
\end{eqnarray}

Finally, we want to make some comments about the critical
dimension of our model. A natural way to find it is to consider
the corresponding BRST anomaly as in the standard string theory
\cite{KO}. This is however a complicated task
due to the infinite number of generators (\ref{new}), but
fortunately, there is no need to solve it completely. The reason
is very simple and to explain it we need only the basic formulas
of the BRST quantization procedure: For a gauge algebra with
generators $J_i$ and structure constants ${f_{ij}}^k$ the BRST
charge is \cite{GSW}
\begin{equation}
Q = c^i J_i - {1\over 2}{f_{ij}}^k c^i c^j b_k  \label{brst}
\end{equation}
where $\{ c^i, b_i\}$ are the ghost-antighost pairs corresponding
to the generators $J_i$. The total symmetry generators (with
the ghost contributions) are
\begin{equation}
J^{tot}_i = J_i + J^{gh}_i = [Q, b_i]_+ =
J_i - {f_{ij}}^k c^j b_k                   \label{gen}
\end{equation}
The generators $J_i^{gh} = -{f_{ij}}^k c^j b_k$
satisfy the same algebra as $J_i$ and therefore the central
extension for them up to a multiplicative constant is the same as
for $J_i$. (The central extension in both cases is equal to the
vacuum expectation value of the corresponding commutator.)
For our algebra there is only one unfixed constant in the
possible central extension and so, the cancellation of the anomaly
in one commutator will lead automatically to the cancellation of
the entire anomaly in all commutators. This fact allows us to
consider only one commutator and in what follows we shall
concentrate our attention on the anomaly in
${\cal G}_1$---${\cal G}_1$
commutator (Virasoro subalgebra).

A short note has to be added at this point. It concerns
our choice of the commutator in which we estimate the anomaly.
At a first sight the simplest one is not the commutator we plan to
investigate but the
${\cal G}_0$---${\cal G}_0$ one. However, since ${\cal G}_0$
forms an affine $U(1)$ algebra there are problems with the
definition of the BRST charge. Our experience shows \cite{SZ2} that in
this case a modification of the BRST procedure is needed but this
question is far away from our present goal and we shall
not deal with it here.

Bellow we use the notations
$\{ c^{\cal G}_m , b^{\cal G}_m\},
\{ c^{\cal X}_m , b^{\cal X}_m\} ,
\{ c^{\cal P}_m , b^{\cal P}_m\}$
for the ghost-antighost pairs corresponding to the constraints
${\cal G}_m, {\cal X}_m$ and ${\cal P}_m$ respectively. We use
also the notations $p^\mu_i, x^\mu_i, c^{\cal G}_{m,i},
b^{\cal G}_{m,i}$
and so on for the Fourier components of
$P^\mu, X^\mu, c^{\cal G}_m, b^{\cal G}_m$ and so on.
Using eqs.(\ref{px-px}--\ref{pp-xx}) and (\ref{brst}) it is easy to
obtain the BRST charge for our algebra and to select the terms
which determine ${\cal G}^{gh}_1$. From them only the following
ones contribute to the Virasoro anomaly
\begin{eqnarray}
& & \sum_{m=0}^\infty \int
[ m c^{\cal G}_m c^{{\cal G}(1)}_1
- c^{\cal G}_1 c^{{\cal G}(1)}_m  ] b^{\cal G}_m  \nonumber \\
&&+ \sum_{m=0}^\infty \int
[ (2m+1) c^{\cal P}_m c^{{\cal G}(1)}_1
+ c^{\cal G}_1 c^{{\cal P}(1)}_m  ] b^{\cal P}_m  \nonumber \\
&&+ \sum_{m=0}^\infty \int
[ (2m-1) c^{\cal X}_m c^{{\cal G}(1)}_1
+ c^{\cal G}_1 c^{{\cal X}(1)}_m  ] b^{\cal X}_m  \nonumber
\end{eqnarray}
According to (\ref{gen})
\begin{eqnarray}
{\cal G}^{gh}_{1,i} &= \cdots & +
\sum_{m=0}^\infty \sum_{j=-\infty}^\infty
(j-mi)c^{\cal G}_{m,-j} b^{\cal G}_{m,i+j}     \nonumber \\
&& + \sum_{m=0}^\infty \sum_{j=-\infty}^\infty
(j-(2m+1)i)c^{\cal P}_{m,-j} b^{\cal P}_{m,i+j}  \nonumber \\
&& + \sum_{m=0}^\infty \sum_{j=-\infty}^\infty
(j-(2m-1)i)c^{\cal X}_{m,-j} b^{\cal X}_{m,i+j} \nonumber
\end{eqnarray}
where dots ($\cdots$) stands for the terms which do not contribute
to the anomaly. Using a standard ghost vacuum for the anomaly we get
\begin{eqnarray}
<0\vert [ {\cal G}^{gh}_{1,i}, {\cal G}^{gh}_{1,i'} ]\vert 0 > &=&
-\delta_{i+i'}\sum_{m=0}^\infty \left[ (9 m^2 +5m +{5\over 2}) i^3
- {1\over 2} i \right]   \nonumber \\
&=&
- \delta_{i+i'} \left[ \sum_{m=1}^\infty
\left[(9 m^2 +5m +{5\over 2}) i^3 - {1\over 2} i \right]
+ {5\over 2} i^3 - {1\over 2}i \right]
 \nonumber \\
&=&
- \delta_{i+i'} \left[ {5\over 6} i^3 - {1\over 4}i \right]
\label{ghan}
\end{eqnarray}
To obtain the last line in the above equation we use the
$\zeta$-function regularization of the sums. The estimation of
the matter field anomaly is simpler.
Skipping the details, the answer we find is
\begin{equation}
<0\vert [ {\cal G}_{1,i}, {\cal G}_{1,i'} ]\vert 0 > =
 \delta_{i+i'}\left[ {D\over 6}i^3 - {D\over 6}i - 2i\beta\right]
\label{matan}
\end{equation}
where $\beta$ is the normal ordering ambiguity constant in ${\cal
G}_1$. From eqs.(\ref{ghan}, \ref{matan}) we see that the model
is anomaly free if
\begin{equation}
D = 5,\;\;\;\;\beta = -{7\over 24} \label{final}
\end{equation}

The obtained critical dimension is ``more physical'' compared to
that of the pure $W_\infty$ string --- The contribution of the
$DOP(S^1)$ ghosts to the anomaly (\ref{ghan}) is
$$
- \delta_{i+i'} \sum_{m=0}^\infty \left[ (m^2 + m +{1\over 6}) i^3
- {1\over 6} i \right].
$$
Proceeding as above one can find that the critical dimension of
the pure $W_{1+\infty}$ string (model with only $DOP(S^1)$
constraints) is
$$ D = 0$$
and that of the $W_\infty$ string (model without spin one generator
${\cal G}_0$) is
$$ D = -1.$$
The latter value differs from that obtained earlier \cite{PRS}
for the $W_\infty$ string which is $D=-2$. The difference is due to
the realization of the (matter field) generators ${\cal G}$ used here.
As a consequence of this realization if we consider only the pure
Virasoro subalgebra we shall get $D=13$ instead of $D=26$. A final
note: All remarks in \cite{PRS} concerning the arbitrariness in the
determination of $D$ are also valid here.
\vskip 0.2cm
{\bf  \hskip 3cm  Acknowledgements}
\vskip 0.2cm
One of us (M.S.) would like to thank Prof. A. Salam,
the International Atomic Energy Agency, and the United Nations
Educational, Scientific and Cultural Organization
for the hospitality
in the International Centre for Theoretical Physics, Trieste,
where the final part of this article has been done.


\begin{thebibliography}{99}
%
\bibitem{SZ}
M. Stoilov and R. Zaikov,
{\it Lett. Math. Phys.} {\bf 27} (1993) 155.
%
\bibitem{Feig}
B. L. Feigin
{\it Usp. Mat. Nauk} {\bf 35} (1988) 157.
%
\bibitem{Mieg}
See, e.g., Jean Thierry-Mieg,
{\it Introduction to Kac-Moody algebras},
preprint DAMTP 88-17 (1988)
%
\bibitem{Rad}
A. O. Radul
{\it Funct. Anal. Appl.} {\bf 25} (1991) 33.
%
\bibitem{KO}
M. Kato and K. Ogawa, {\it Nucl. Phys. } {\bf B212} (1983) 445.
%
\bibitem{GSW}
See, e.g., M. Green, J. Schwartz and E. Witten,
{\it Superstring Theory } {\bf vol~I}, Cambridge University
Press, 1987.
%
\bibitem{SZ2}
M. Stoilov and R. Zaikov,
{\it Mod. Phys. Lett. } {\bf A8} (1993) 2687.
%
\bibitem{PRS}
P. N. Pope, L. J. Romans and X. Shen,
{\it Phys. Lett. } {\bf B254} (1991) 401.
%
\end{thebibliography}
\end{document}